\documentclass[aps,prl,reprint,showpacs,superscriptaddress]{revtex4-2}

\usepackage{amsmath}
\usepackage{graphicx}
\usepackage{lmodern}
\usepackage{amsmath}
\usepackage{color}
\usepackage{amssymb}
\usepackage{bm}
\usepackage{dsfont}
\usepackage{braket}
\usepackage{mathtools}
\usepackage{afterpage}

\newcommand{\br}{\bm{r}}
 
\newcommand{\bR}{\bm{R}}
\newcommand{\bx}{\bm{x}}

\newcommand{\bk}{\bm{k}}

\newcommand{\bG}{\bm{G}}

\newcommand{\be}{\begin{equation}}
\newcommand{\ee}{\end{equation}}

\usepackage[dvipsnames]{xcolor}
\usepackage[colorlinks=true]{hyperref}
\hypersetup{
    colorlinks=true,
    linkcolor=blue,
    citecolor=blue,
    urlcolor=blue
} 

\makeatletter
\def\maketitle{
\@author@finish
\title@column\titleblock@produce
\suppressfloats[t]}
\makeatother

\begin{document}
\title{Duality Between Twist-Angle Disorder and Non-Hermitian Disorder}

\author{Yi-Ming Wu}
\email{yimwu@zju.edu.cn}
\affiliation{Institute for Advanced Study in Physics, Zhejiang University, Hangzhou 310027, China}

\author{Nicole S. Ticea}
\email{nticea@stanford.edu}
\affiliation{Department of Applied Physics, Stanford University, Stanford, CA 94305, USA}
\affiliation{Google Quantum AI, Goleta, CA, 93111, USA}

\date{\today}

\begin{abstract}
  We propose and study a model of moiré heterobilayer systems where the twist angle remains uniform and well defined only within a finite range, but deforms randomly at larger distances. The local twist angle is then subject to a certain probability distribution that  penalizes large fluctuations around its mean value. Within the framework of the replica trick, we show that disorder averaging produces a nontrivial correlation with alternating sign between replicas, maximized only close the boundary of each local domain and along certain directions. We show that for the low-energy moiré bands, exactly the same correlation can be generated in a dual model where fermions are coupled to non-Hermitian disorder, which can evade Anderson localization dynamically. This duality provides a new perspective to investigate twist-angle disorder in moiré systems,  and unveils some of the essential differences between twist-angle disorder and conventional disorder.
\end{abstract}
\maketitle

{\it Introduction.}-- Twist-angle disorder is ubiquitous in moiré systems, including twisted bilayer and trilayer graphene and twisted homo- and heterobilayers of transition-metal dichalcogenides (TMDs)\cite{Uri2020,PhysRevResearch.3.013153,Kazmierczak2021,SimonTurkel,deJong2022,bathen2025precisetwistangledetermination}. It represents a new form of disorder unique to two-dimensional moiré van der Waals materials: even when the constituent monolayers are intrinsically clean, local strain, stress, and structural inhomogeneity can produce spatial variations in the twist angle. Since in 2D the electron states may have already been modified substantially by disorder at the single-particle level, 
understanding this disorder is essential for developing a complete picture of correlated phases in moiré materials, including superconductivity\cite{Cao2018,Yankowitz2019,Lu2019,Park2021,Xia2025,Guo2025}, quantum anomalous Hall states\cite{Serlin2020,Sharpe2019,Nuckolls2020,Wu2021,Choi2021,Chen2020,LiQAH2021,Zhang2019,LiuDai2021}, and fractional Chern (topological) insulators\cite{Xie2021,Spanton2018,Cai2023,Zeng2023,Park2023,Xu2023,LiFCI2021,Crepel2023,Ledwith2020,Repellin2020}.
For example, although conventional $s$-wave superconductivity is relatively robust against nonmagnetic disorder, unconventional pairing states proposed for twisted bilayer graphene and twisted bilayer WSe$_2$ and MoTe$_2$ may be particularly sensitive to twist-angle inhomogeneity\cite{Anderson1959,Oh2021,Isobe2018,Wu2018,XuBalents2018,LiuPairing2018,Kennes2018,Gonzalez2019,Samajdar2020,Chou2021,Wu2023,SchradeFu2024,Chen2026,Xia2025,Guo2025,Uri2020}. Likewise, given the well-known sensitivity of fractional quantum Hall states to disorder\cite{MacDonald1986,Sheng2003,Deng2014,WangDisorder2018,Zhu2019}, the fractional Chern insulating states observed in twisted bilayer MoTe$_2$ may also be fragile against twist-angle disorder\cite{Cai2023,Zeng2023,Park2023,Xu2023,Wang2024,Reddy2023,Yu2024,MoralesDuran2024,Jia2024,Redekop2024,PhysRevB.109.115111,PhysRevLett.133.186602,3crp-gb3d}. Clarifying the nature and consequences of this distinct form of disorder is therefore crucial for understanding the stability and phenomenology of correlated phases across moiré platforms.

Unlike conventional impurity-induced disorder, which is typically modeled as a random potential coupled to the electron density, twist-angle disorder is considerably more difficult to describe, although some recent studies use Anderson model to mimic the disorder effects in TBG\cite{,PhysRevLett.134.126301,9lc1-9m8t}. More realistic attempts have mapped the twist-angle disorder in TBG to problems involving spatially random Fermi velocities\cite{PhysRevResearch.2.043416} and random gauges fields\cite{c6rt-6qg1}, or incorporated a nonuniform lattice distortion in the continuum model\cite{PhysRevB.105.245408}. Large-scale tight-binding simulations have also been employed to investigate its effects on electronic transport\cite{PhysRevResearch.2.023325,ciepielewski2024transporteffectstwistangledisorder}. Despite these efforts, a simple and physically relevant model that captures the essential nature of twist-angle disorder is still lacking.

Here in this Letter, we study a model moiré bilayer system in which the twist angles are arranged to form local domains with randomly distributed local twist angles. We show that under disorder averaging (valid for large-size systems where self-averaging is meaningful), the twist-angle disorder in the low-energy moiré bands admits a dual description in which electrons close to the domain boundaries are coupled to {\it non-Hermitian} disorder potentials. Like the conventional case of Hermitian disorder, non-Hermitian disorder does localize electron wavefunctions. However, unlike its Hermitian counterpart, non-Hermitian disorder allows for jumpy events between localized states, dynamically avoiding Anderson localization\cite{z9m1-3mwb,Weidemann2021,Longhi2023}. This is consistent with numerical results starting from a more microscopic model showing enhanced diffusion by twist-angle disorder\cite{paper2}. In identifying the duality between twist-angle disorder and non-Hermitian disorder we have proposed a bridge between seemingly unrelated subjects, in service of the greater goal of understanding the complete phase diagram of moiré systems. 

\begin{figure}
    \centering
    \includegraphics[width=\linewidth]{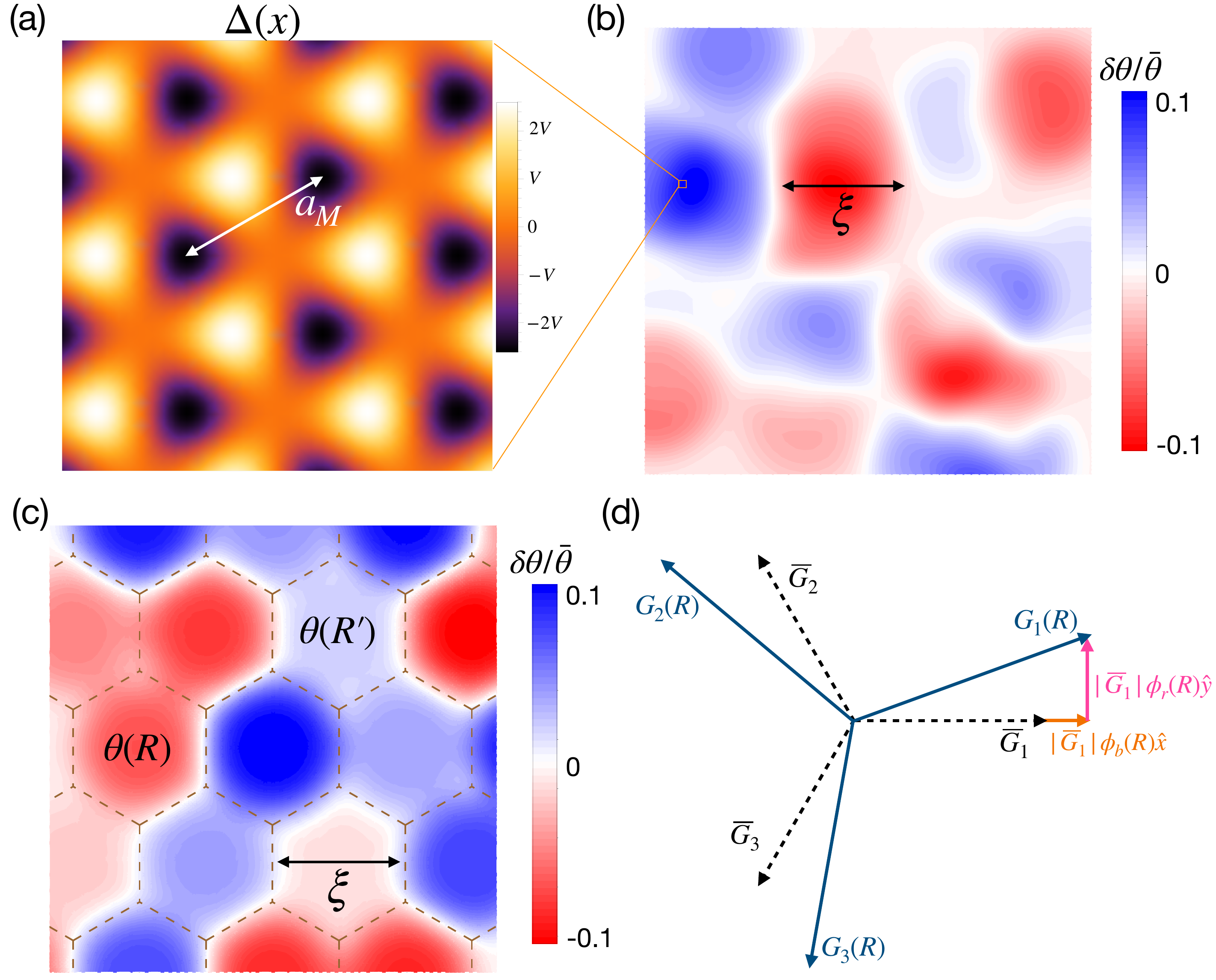} 
    \caption{(a) Moiré potential $\Delta(\bx)$ with constant twist angle, defined in Eq.\eqref{eq:H}, with the parameter $\varphi\approx \pi/2$. The periodicity is set by the moiré unit cell length $a_M$. 
    (b) In the presence of twist-angle disorder, there are domains of uniform twist angle whose length is set by $\xi\gg a_M$. (c) Our model of twist-angle disorder. The system is divided into  domains of the same size $\xi$, labeled by $\bR$. Within each supercell the twist angle is approximated as a constant, which permits a set of local lattice vectors $\bm{a}_j(\bR)$, and hence $\bG_j(\bR)$. (d) We parametrize the variations of $\bG_j(\bR)$  by two scalar fields $\phi_r(\bR)$ and $\phi_b(\bR)$, see Eq.\eqref{eq:G1R}
    }
    \label{fig:model}
\end{figure}

{\it Model and method.}-- For convenience of discussion, we consider the simplest moiré system: twisted bilayer WSe$_2$/MoSe$_2$\cite{PhysRevLett.121.026402}. Generalization of our results to other moiré systems is straightforward. 
Here both layers have almost the same lattice constant but their valence bands are located at different energies.
Moreover, these TMDs possess a large, valley-contrasting spin-orbit coupling (SOC) in the valence bands, 
effectively giving rise to spin-valley locking and reducing the degrees of freedom for the moiré band structures. 
It is therefore possible for the Fermi level to cross the valence band of one layer while remaining within the gap of the other layer; as a result the low-energy moiré bands for each valley in the `clean' limit (with no angle disorder) can be effectively described by the following continuum limit Hamiltonian,
\begin{equation}
     \begin{aligned}
         H&=-\bk^2/(2M)+\Delta(\br),\\
     \Delta(\bx)&=\frac{V}{2}e^{i\varphi}\sum_{j=1}^3 e^{i\bm{G}_j\cdot \bx}+h.c.,
     \end{aligned}\label{eq:H}
 \end{equation} 
where $M$ is the valence band effective mass, $V>0$ is the potential strength, $\varphi$ is a phase factor, and $\bm{G}_{j}$ are three reciprocal lattice vectors related to each other by $\pm\frac{2\pi}{3}$ rotation.
Note that the twist angle $\theta$ enters $H_\sigma$ through $\bm{G}_j$ since its magnitude is given by $|\bm{G}_j|=4\pi/(\sqrt{3}a_M)$ and $a_M\approx a_0/\theta$ is the moiré lattice constant ($a_0$ is the monolayer atomic lattice constant).
By fitting to DFT calculations it is found that $\varphi\approx\pi/2$. Thus, we will hereafter fix $\varphi$ to be $\pi/2$, in which case the minima and maxima of $\Delta(\br)$ are opposite to each other, forming a triangular moiré lattice, as shown in Fig.\ref{fig:model}(a).

In a realistic system, the twist angle $\theta$ may become spatially nonuniform and the system may tend to develop domains (of typical size $\xi$) across which $\theta=\bar\theta+\delta\theta$ varies randomly, as demonstrated in Fig.\ref{fig:model}(b). Although the realistic domains are amorphous and irregular, the essential physics shall not depend on the explicit arrangement of the domains. Thus, for the ease of theoretical analysis, we propose a model in which the domains of size $\xi$ are arranged to form a regular pattern as in Fig.\ref{fig:model}(c). We denote the local twist angle by $\theta(\bR)$ with $\bR$ the position of each domain center. $\theta(\bR)$ varies randomly above an averaged value $\bar\theta$ as $\bR$ changes. 
Note that the legitimacy of the local twist angle description requires $\xi$ to be much larger than $a_M$. Furthermore, to validate the self-averaging in our approach below, we also assume that the system size, denoted by $L$, is much larger than $\xi$. These considerations naturally give rise to a hierarchy,
\begin{equation}
    a_M\ll\xi\ll L.\label{eq:hierarchy}
\end{equation}

As a result of the spatially varying $\theta(\bR)$, the moiré lattice vectors $\bm{a}_j$ ($\sim a_M$) and the reciprocal lattice vectors $\bm{G}_j$ all become domain dependent, as they are related by $\bm{G}_i(\bR)\cdot \bm{a}_j(\bR)=2\pi\delta_{i,j}$ where $i,j=1,2$ label the two independent lattice vectors. 
We denote the spatially averaged reciprocal vectors as $\overline{\bm{G}}_j$, which are given by $\overline\theta$. Without loss of generality we assume $\overline{\bm{G}}_1$ is along $\hat x$ direction, and parametrize its variation as [see also Fig.\ref{fig:model}(d)]
\begin{equation}
    \bm{G}_1(\bR)=|\overline{\bm{G}_1}|\left[\hat x+\phi_b(\bR)\hat x + \phi_r(\bR)\hat y\right],
     \label{eq:G1R}
\end{equation}
where $\phi_{r,b}(\bR)$ are two {\it small} random scalar fields, varying slowly at the scale of $\bR$ ($\sim\xi$). 
Physically, $\phi_b(\bR)$ indicates the `breathing' deformation of the moiré lattice, while $\phi_r(\bR)$ implies the `rotating' deformation. Note in order to define the local $\theta(\bR)$, we have preserved the $C_3$ rotation symmetry in each domain, so that $\phi_r(\bR)$ and $\phi_r(\bR)$ are the only two independent allowed deformations. The other two vectors $\bm{G}_{2,3}(\bR)$ are obtained by $\pm2\pi/3$ rotations from $\bG_1$.
Now we are in a position to state the key ingredient of our theory: the 
probability distribution function (pdf) for $\phi_{\nu=b,r}$,
\begin{equation}
    \mathcal{P}[\phi_{\nu}] = \exp\left\{-\frac{\kappa_\nu}{2\pi}\int d^2\bR \Big[(\nabla_{\bR} \phi_{\nu}(\bR))^2 + m_\nu^2 \phi_{\nu}^2(\bR) \Big] \right\}.\label{eq:phi_action}
\end{equation}
Here $\kappa_\nu$ (dimensionless) is an effective stiffness that penalizes fast fluctuations of $\phi_\nu$, and $m_\nu$ is a mass term that guarantees $\overline\phi_\nu=0$. 
Note that $\phi_\nu$ is introduced as some quenched disorder field; thus it has no dynamics.

With the angle disorder encoded in $\bG_j(\bR)$, the moiré potential $\Delta(\bx)$ introduced in Eq.\eqref{eq:H} is  modified into $\Delta_{\bR}(\br)$ with the identification $\bx=\bR+\br$ ($\br$ is now measured from each domain center so that $|\br|<\xi$). Explicitly, 
\begin{equation}
        \Delta_{\bR}(\br) =\frac{V}{2}e^{i\varphi}\sum_{j=1}^3e^{i\overline{\bm{G}}_j\cdot \br} \mathcal{V}_j(\bR,\br)+ \text{h.c.}\label{eq:disorderpotential1}
\end{equation}    
where we have defined the vertex function 
\begin{equation}
    \mathcal{V}_j(\bR,\br)=e^{i[\beta_{b,j}(\br)\phi_b(\bR)+\beta_{ r,j}(\br)\phi_r(\bR)]}\label{eq:vertexF}
\end{equation}
and the  `frequencies' are $\beta_{b,j}(\br)=\frac{4\pi|\br|}{\sqrt{3}\overline{a_M}}\cos[\theta_{\br}-\tfrac{2\pi}{3}(j-1)]$ and $\beta_{r,j}(\br)=\frac{4\pi|\br|}{\sqrt{3}\overline{a_M}}\sin[\theta_{\br}-\tfrac{2\pi}{3}(j-1)]$.
Here $\theta_{\br}$ is the angle between $\br$ and $\hat x$.  
The problem of free electrons propagating in such a disordered moiré potential is then given by the full action $ S=S_0+S_\text{pot}$, where 
\begin{equation}
   \begin{aligned}
       S_0&=\int d\tau\int d^2\bm{x} {\psi}^\dagger(\tau,\bm{x})\left(\partial_\tau+\frac{\nabla_{\bm{x}}^2}{2M}\right)\psi(\tau,\bm{x}),\\
       S_\text{pot}&=\int d\tau \sum_{\bR}\int d^2\br{\psi}^\dagger_{\bR}(\tau,\br){\psi}_{\bR}(\tau,\br)\Delta_{\bR}(\br).
   \end{aligned}\label{eq:S_0}
\end{equation}
We have 
adopted the Euclidean convention with $\tau$ the imaginary time. 
The fermion field is denoted either by $\psi(\tau,\bm{x})$, or equivalently by $\psi_{\bR}(\tau,\br)=\psi(\tau,\bR+\br)$, as long as we identify $\sum_R\to \frac{1}{\xi^2}\int d^2\bR$.

Having established the disordered moiré potential and its coupling to electrons, we can investigate the limit $\xi\ll L$ [see Eq.\eqref{eq:hierarchy}], where the disorder averaging is meaningful. 
Here we use the replica trick; instead of evaluating the disorder average of $\ln Z$, we will be interested in evaluating the disorder average of the replicated $Z^N$, $N$ being the number of the replicas. $Z$ can be any generating functional of interest.
To see how this works, let's introduce a compact notation
\begin{equation}
    \alpha_j(x)=\frac{V}{2}e^{i\varphi}e^{i\overline{\bm{G}}_j\cdot \br} \sum_{a=1}^N{\psi}^{a \dagger}_{\bR}(\tau,\br){\psi}_{\bR}^a(\tau,\br),
\end{equation}
where $a$ is the replica index, and $x=(\tau,\bm{x})=(\tau,\bR,\br)$ is a compact representation of the space-time coordinates. 
Defining the disorder average as $\overline{(\cdots)}=\int\prod_{\nu=b,r}\mathcal{D}[\phi_\nu]\mathcal{P}[\phi_\nu](\cdots)$, we have
\begin{equation}
        \overline{e^{-S_\text{pot}}}=\overline{\exp\left\{-\int \frac{dx}{2\xi^2}\sum_j\alpha_j(x)\mathcal{V}_j(\bR,\br)+ \text{h.c.} \right\}}\label{eq:disorderav1}
\end{equation}
We remark that the twist-angle disorder average is essentially averaging the exponential of exponentials of $\phi_\nu$ fields.

{\it Disorder average.--} To evaluate Eq.\eqref{eq:disorderav1} in a controllable manner, we assume the disorder is weak and use the method of cumulant expansion.
Up to second order in the cumulants, we have $\overline{e^{-S_\text{pot}}}\approx e^{-S_\text{dis}}$ where $S_\text{dis}=\overline{S_\text{pot}}-\frac{1}{2}\left(\overline{S_\text{pot}^2}-\overline{S_\text{pot}}^2\right)\equiv S_\text{dis}^{(1)}+ S_\text{dis}^{(2)}$. The effective action, after disorder averaging, is
\begin{equation}
S_\text{eff}=S_0+S_\text{dis}^{(1)}+S_\text{dis}^{(2)}
\end{equation}
Note that $S_0$ here also contains $N$ replicas. Under this approximation, we still need to compute the disorder average for $\mathcal{V}_j(\bR,\br)$ subject to the pdf in Eq.\eqref{eq:phi_action}. To avoid unphysical UV divergences from self-contractions of $\phi_\nu(\bR)$, we need to ``normal order'' the fields before averaging. For the vertex function introduced in Eq.\eqref{eq:vertexF}, normal ordering is simply $:\mathcal{V}_\mu(\bR,\br):=\mathcal{V}_\mu(\bR,\br)/\overline{\mathcal{V}_\mu(\bR,\br)}$ so that $\overline{:\mathcal{V}_\mu(\bR,\br):}=1$. We obtain
\begin{equation}
    S_\text{dis}^{(1)}=\overline{S_\text{pot}}=\int\frac{dx}{2\xi^2}\sum_{j=1}^3\alpha_j(x)+h.c.
\end{equation}
Apparently, the first-order effect is nothing but to just treat fermions in an averaged, uniform moiré potential determined by the twist angle $\overline{\theta}$.

Moving on to the next order term, $S_\text{dis}^{(2)}$, we need 
the following correlation functions 
\begin{equation}
     \begin{aligned}
         &V^{(2,s)}_{jj'}(\bR,\br;\bR',\br')\\
         &=\overline{:e^{\pm i \sum_\nu\beta_{\nu,j}(\br)\phi_\nu(\bR)}::e^{\mp is\sum_{\nu'}\beta_{\nu',j'}(\br')\phi_{\nu'}(\bR')}:}\\
         &= e^{s\sum_{\nu,\nu'}\beta_{\nu,j}(\br)\beta_{\nu',j'}(\br')C_{\nu\nu'}(\bR-\bR')}, ~~ s=\pm1.\\
     \end{aligned}\label{eq:V1V2}
 \end{equation} 
 and $C_{\nu\nu'}(\bR-\bR')=\overline{\phi_\nu(\bR)\phi_{\nu'}(\bR')}\equiv\frac{\delta_{\nu,\nu'}}{2\kappa_\nu}K_0(m_\nu|\bR-\bR'|)$. $K_0$ is the modified Bessel function of the second kind which defines another length scale $\ell_\nu=m_\nu^{-1}$. Because $\ell_\nu$ measures how far a local twist angle deformation in one domain can propagate, we naturally require $\ell_\nu\gtrsim\xi$. Note that when $\xi\leq|\delta \bR|\ll\ell_\nu$, $C_{\nu\nu}(\delta\bR)\sim\ln\frac{L^2}{\delta\bR^2}$, while for $|\delta\bR|\gg\ell_\nu$ $C_{\nu\nu}(\delta\bR)\sim\exp(-|\delta\bR|/\ell_\nu)$.
It is easy to see that the largest value of $V_{jj'}^{(2,s)}$ occurs when both $\br$ and $\br'$ are close to their domain boundaries, and when $\bR$ and $\bR'$ label adjacent domains. In this case, one can estimate by taking $\kappa_b=\kappa_r\equiv\kappa$ that its maximal value is $V_\text{max}^{(2,s)}\approx(L/\xi)^{\frac{2\pi\xi}{\sqrt{3}\kappa\overline{a_M}}}$.
When either $\br$ or $\br'$ is shifted away from the domain boundary, $V$ becomes exponentially smaller. This observation also defines what we mean by weak disorder: $\kappa_\nu$ has to be large enough such that the largest value of $V_{jj'}^{(2,s)}$ is still not far from $1$---the clean limit obtained from taking $\kappa_\nu\to\infty$. 
In terms of $V_{jj'}^{(2,s)}$, it is straightforward to see
\begin{equation}
    S_\text{dis}^{(2)}=-\frac{V^2}{4\xi^4}\int d\tau d\tau' d\bm{x}d\bm{x}' \rho(\tau,\bm{x})\rho(\tau',\bm{x}')g(\br,\br';\bR-\bR'),\label{eq:Sdis2}
\end{equation}
where $\rho(\tau,\bm{x})=\sum_{a=1}^N{\psi}^{a \dagger}_{\bR}(\tau,\br){\psi}_{\bR}^a(\tau,\br)$ is the replicated fermion density and 
  \begin{equation}
    \begin{aligned}
         g(\br,\br';\bR-\bR')&=\sum_{jj',s=\pm1}s\cos\left(\overline{\bm{G}}_j\cdot \br-s\overline{\bm{G}}_{j'}\cdot\br'\right)\\
         &\times\left(V^{(2,s)}_{jj'}(\bR,\br;\bR',\br')-1\right).\label{eq:grR}
    \end{aligned}
  \end{equation}
In arriving at this expression, we have already used the condition that $\varphi=\pi/2$.
In the clean limit, $V_{jj'}^{(2,s)}=1$ and Eq.\eqref{eq:grR} vanishes as it should. When $\bR$ and $\bR'$ are separated at a distance larger than $\ell_\nu$, $g(\br,\br;\bR-\bR')$ is also negligible. 

\begin{figure}
    \centering
    \includegraphics[width=\linewidth]{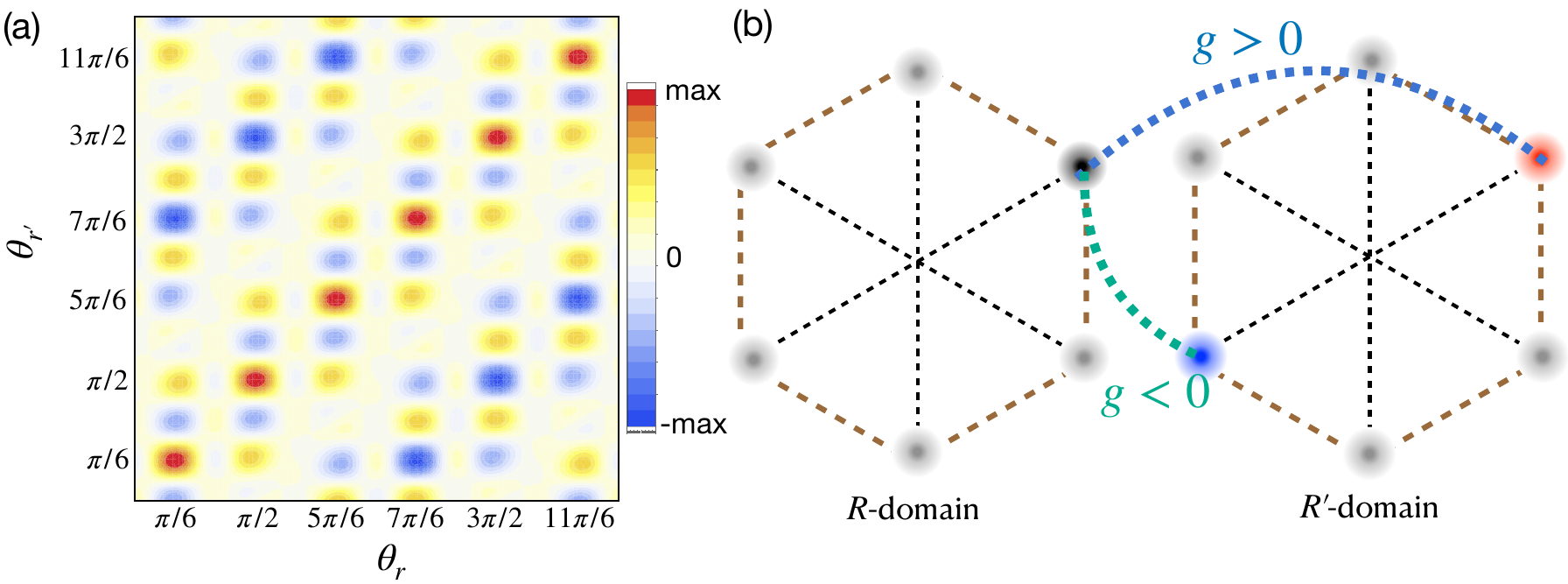}      
    \caption{(a) Plot of $g(\br,\br';\bR-\bR')$ as a function of $\theta_{\br}$ and $\theta_{\br'}$ when $|\bR-\bR'|$ is fixed and $|\br|=|\br'|$ is set to be $\sim\xi/2$. (b) The peak (maximum and minimum) positions in (a) implies only six `hot spot' at each domain boundary are effectively correlated , and correlations between different domains are such that when $\theta_{\br}=\theta_{\br'}$, $g>0$, and when $\theta_{\br}=\theta_{\br'}\pm\pi$, $g<0$, as implied from (a).}
    \label{fig:gfactor}
\end{figure}
\begin{figure}
    \includegraphics[width=\linewidth]{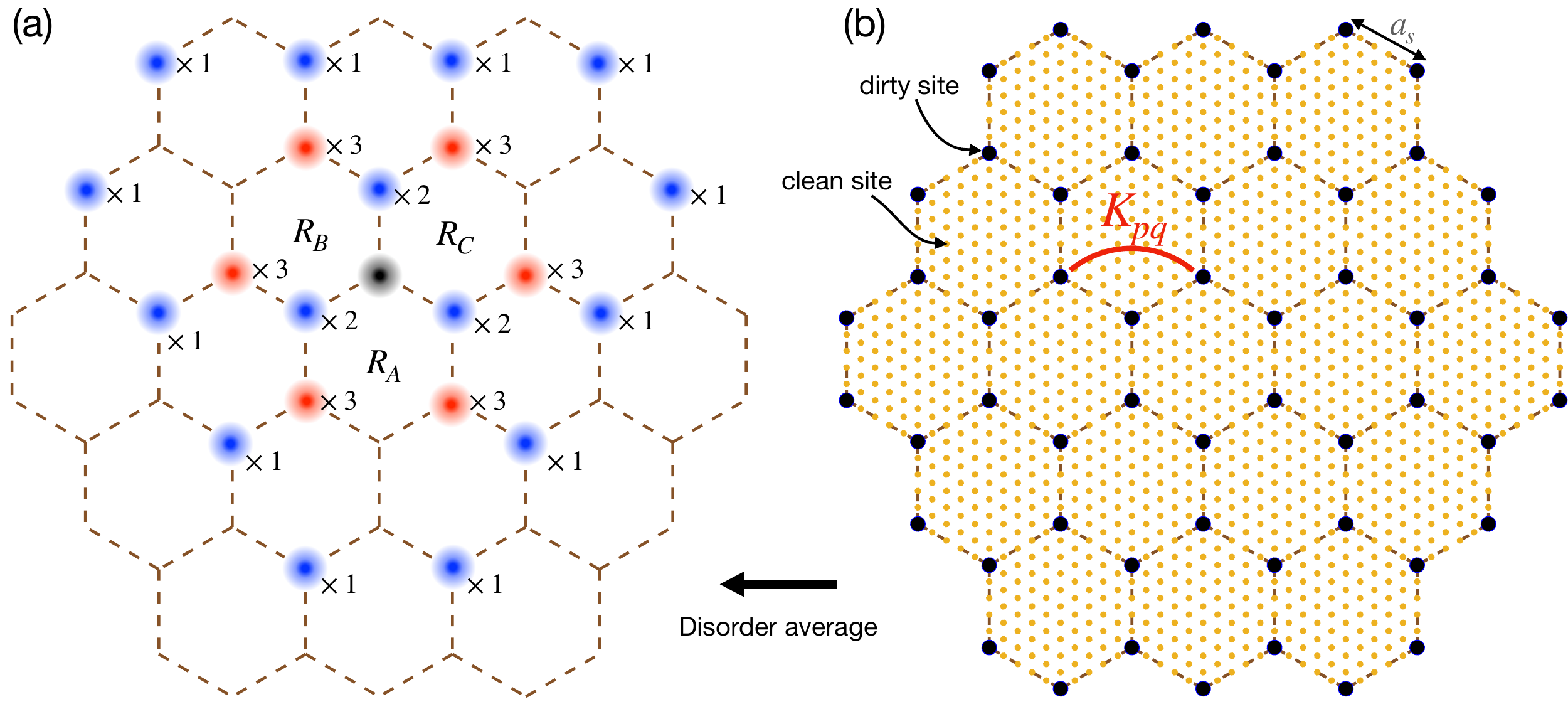}\caption{(a) Pattern of correlations with the central spot (black) induced by the twist-angle disorder when only the nearest neighbor domains are correlated ($\ell_\nu\sim\xi$). Here the black spot correlates with the red ones positively, and with the blue ones negatively. The indicated integer multiples arise from the fact that we can view the black spot as the corners of $\bR_A$, $\bR_B$ and $\bR_C$ domains. (b) In the tight-binding description, the correlation in (a) can be equivalently induced by adding random onsite potentials to the corner sites (black dots), if the potential obey the same correlation $\overline{v_p^*v_q}=K_{pq}$. }\label{fig:correlationPattern}
\end{figure}

In Fig.\ref{fig:gfactor} (a) we present such a plot in the $(\theta_{\br},\theta_{\br'})$ plane with a fixed $\bR-\bR'$ and $|\br|=|\br'|=\xi/2$, i.e. both $\br$ and $\br'$ are placed at their domain boundaries such that $V_{jj}^{(2,s)}$ and hence $g(\br,\br';\bR-\bR')$ obtain their maximum values. Despite having a complicated analytical expression, $g(\br,\br';\delta\bR)$ shows a regular pattern in the $(\theta_{\br},\theta_{\br'})$ plane. 
In particular, $g(\br,\br';\delta\bR)$ reaches its maximum value when $\theta_{\br}=\theta_{\br'}=\frac{\pi}{6}+\frac{n\pi}{3}$ with $n\in \mathbb{Z}$, and it reaches its minimum value when the condition $\theta_{\br}=\theta_{\br'}-\pi=\frac{\pi}{6}+\frac{n\pi}{3}$ is met.
These six global maxima and minima indicate the six discrete spatial directions within each domain along which $g$ can reach its largest value, as illustrated in Fig.\ref{fig:gfactor}(b). 
For each domain, one identifies six intersecting regions between the discrete maximal directions and the domain boundary, marked with gray spots. 
For every such small region on the boundary  of the $\bR$-domain [say the black spot in Fig.\ref{fig:gfactor}(b)], if a small boundary region of $\bR'$-domain aligns in the same direction [red spot in Fig.\ref{fig:gfactor}(b)], these two regions are positively correlated; in contrast if a small boundary region of $\bR'$-domain is aligning in the opposite direction [blue spot in Fig.\ref{fig:gfactor}(b)], these two regions are negatively correlated. Note by considering the correlations only among these boundary regions, the translation symmetry is effectively restored. 
It is in principle possible to continue the above analysis for the local minima and maxima shown as the lighter blue and lighter orange regions in Fig.\ref{fig:gfactor}(a).

{\it Dual description from non-Hermitian disorder.--}
Having identified the form of the correlation induced by twist-angle disorder, we are now in a position to state the dual picture in which the same correlation is obtained by performing disorder averaging for some random impurity coupled to the fermion density.
Formally, this is expressed as 
\begin{equation}
    \overline{Z^N[\psi^\dagger,\psi,\phi_\nu]}=\overline{Z^N_\text{dual}[\psi^\dagger,\psi,v]}
\end{equation}
Namely, the partition function (or generating functional) after averaging the twist-angle disorder is identical to the disorder averaging of a dual problem in which the fermions are coupled to some non-Hermitian disorder $v$. 

To better demonstrate this, we can focus on the first few moiré bands, which are of particular interest  not only because they are experimentally accessible by tuning the gate voltage, but also because they can be fairly easily described by a tight-binding model on the triangular moiré lattice\cite{PhysRevLett.121.026402}. 
In this real space picture, we can choose to focus only on the corner regions since they are mostly correlated based on our analysis of $g(\br,\br';\bR,\bR')$ above. 
How they are correlated is explicitly shown in Fig.\ref{fig:correlationPattern}(a). Let's take a particular corner region, say the black spot, as an example. Depending on how large $\kappa_\nu$ is, it can include multiple lattice sites, or just a single lattice site (the simplest case when $\kappa_\nu$ is large enough). These corner sites can be equivalently viewed as belonging to the boundary of the $\bR_{A}$, $\bR_B$, or $\bR_C$ domains. 
One then identifies the neighboring corner sites that are either positively or negatively correlated with the black spot, according to the rule in Fig.\ref{fig:gfactor}(b). Restricting to the case when only nearest neighbor domains are correlated (i.e. $\ell_\nu\sim\xi$) 
results in a more regularized pattern (presented in Fig.\ref{fig:correlationPattern}(a)), where the integer multiples come from addition of the three domains. For the lattice model, we can write the disorder-induced part as $S_\text{dis}^{(2)}=\sum_{p,q\in \text{Corners}} \frac{1}{2}n^r_p K_{pq}n^r_q$,
supported only on the black corners in Fig.\ref{fig:correlationPattern}(b). Here $n^r_p$ and $n^r_q$ are abbreviations for the replicated density operators integrated over imaginary time $\tau$. The kernel matrix $K_{pq}$ contains the correlation pattern in Fig.\ref{fig:correlationPattern}(a), and is given by $ K_{pq}=2g$ if $|\br_{p}-\br_{q}|=a_s$; $ K_{pq}=-3g$ if $|\br_{p}-\br_{q}|=\sqrt{3}a_s$; and $ K_{pq}=g$ if $|\br_{p}-\br_{q}|=\sqrt{7}a_s$ or $\sqrt{13}a_s$.
We have used $g$ to denote the maximum value of $\frac{V^2}{2}g(\br,\br';\bR-\bR')$ and $a_s\sim \xi$ is the length of each domain edge.

The essential feature of the correlation $K_{pq}$ is that as a symmetric matrix it is not positive definite. Because of this, it is not possible to invoke in the dual picture only real disorder. To be precise, the dual lattice model is described by
\begin{equation}
    H_\text{dual}=-t\sum_{\braket{ij}}c_i^\dagger c_j+\sum_{p\in \text{Corners}} v_p n_p,\label{eq:dualH}
\end{equation}
where $\braket{ij}$ means nearest neighbor and $v_p$ is a {\it complex} disorder potential at the dirty site $p$ subject to $\overline{v_p^*v_q}=K_{pq}$. This Hamiltonian thus becomes non-Hermitian. 
To see how to choose a $v_p$ giving the same correlation $K_{pq}$, it is useful to diagonalize the kernel, $K^{-1}=U^T \begin{pmatrix}
    \lambda_+^{-1} & 0\\
    0 & \lambda_-^{-1}
\end{pmatrix}U$, where $\lambda_+$ ($\lambda_-$) contains all the positive (negative) eigenvalues. Let $\xi_+$ and $\xi_-$ be two real random vectors whose probability distribution functions are $P[\xi_\pm]=\exp[-\frac{1}{2}\xi_{\pm}^T\lambda_\pm^{-1}\xi_{\pm}]$. Then the onsite random potentials are given by
\begin{equation}
    v=U^T\begin{pmatrix}
        i\xi_+\\
        \xi_-
    \end{pmatrix}
\end{equation}
Eq.\eqref{eq:dualH} with the random potentials determined this way provides a dual picture for the twist-angle disorder problem.

{\it Discussions.--} We have shown in this Letter that even weak twist-angle disorder in a simple moiré system can result in an unconventional, sign-alternating correlation after disorder averaging, in stark contrast to conventional Anderson disorder. The duality to non-Hermitian disorder we identified reflects the non-triviality of the twist-twist angle disorder. Generalizing our model to more realistic systems with amorphous domains is also possible. According to Eq.\eqref{eq:grR} and Fig.\ref{fig:gfactor}, one needs to first identify the six distinct directions emanating from the domain center, and then locate the intersections between these directions with the domain boundary. The correlation patterns among these boundary regions can then be identified as in our model, although the amorphous shape brings about additional complexities in determining the values of the correlations. Despite differences in the details, the long-wavelength physics should belong to the same universality class, regardless of the shape of the domains. The non-perturbative consequences of this disorder---and the interplay between interactions---can be analyzed through the dual picture, which we leave for future studies. 

{\it Acknowledgments. --} We would like to thank Fengcheng Wu, Trithep Devakul, Sri Raghu, Pavel Nosov, Akshat Pandey, and Yi Huang for inspiring discussions. Y.M.W. is supported by a startup fund at Zhejiang University. N.S.T. is funded by Google Quantum AI.

   \bibliography{tbg}
\end{document}